\documentclass[utf8]{frontiersSCNS} 

\usepackage{url,hyperref,lineno, microtype, subcaption}
\usepackage[onehalfspacing]{setspace}
\usepackage{sidecap}
\usepackage{epstopdf} 

\hypersetup{colorlinks=true, linkcolor=red, citecolor = blue, filecolor=cyan,  urlcolor=magenta,}

\newcommand{\aap}{    {\it Astron. Astrophys.}}

\newcommand{\apj}{    {\it Astrophys. J.}}
\newcommand{\apjl}{   {\it Astrophys. J. Lett.}}

\newcommand{\grl}{    {\it Geophys. Res. Lett.}}

\newcommand{\nat}{    {\it Nature}}

\newcommand{\solphys}{{\it Solar Phys.}}
 
\newcommand{\ssr}{    {\it Space Sci. Rev.}} 
\chardef\us=`\_

\def\keyFont{\fontsize{8}{11}\helveticabold }
\def\firstAuthorLast{P.~Vemareddy} 
\def\Authors{P.~Vemareddy\,$^{1,*}$ }  
\def\Address{$^{1}$Indian Institute of Astrophysics, II Block, Koramangala, Bengaluru-560 034, India 
}
\def\corrAuthor{P.~Vemareddy}

\def\corrEmail{vemareddy@iiap.res.in}

\begin{document}
\onecolumn
\firstpage{1}

\title[Magnetic Structure in Successively Erupting Active Regions]{Magnetic Structure in Successively Erupting Active Regions: Comparison of Flare-Ribbons with Quasi-Separatrix Layers } 

\author[\firstAuthorLast ]{\Authors} 
\address{} 
\correspondance{} 

\extraAuth{}

\maketitle 

\begin{abstract}
	This paper studies the magnetic topology of successively erupting active regions (ARs) 11429 and 12371. Employing vector magnetic field observations from Helioseismic and Magnetic Imager, the pre-eruptive magnetic structure is reconstructed by a model of non-linear force-free field (NLFFF). For all the five CMEs from these ARs, the pre-eruptive magnetic structure identifies an inverse-S sigmoid consistent with the coronal plasma tracers in EUV observations. In all the eruption cases, the quasi-separatrix layers (QSLs) of Large Q values are continuously enclosing core field bipolar regions in which inverse-S shaped flare ribbons are observed. These QSLs essentially represent the large connectivity gradients between the  domains of twisted core flux within the inner bipolar region and  the surrounding potential like arcade. It is consistent with the observed field structure largely with the sheared arcade. The QSL maps in the chromosphere are compared with the flare-ribbons observed at the peak time of the flares. The flare ribbons are largely inverse-S shape morphology with their continuity of visibility is missing in the observations. For the CMEs in the AR 12371, the QSLs outline the flare ribbons as a combination of two inverse J-shape sections with their straight parts being separated. These QSLs are typical with the weakly twisted flux rope. Similarly, for the CMEs in the AR 11429, the QSLs are co-spatial with the flare ribbons both in the middle of the PIL and in the hook sections. In the frame work of standard model of eruptions, the observed flare ribbons are the characteristic of the pre-eruptive magnetic structure being sigmoid which is reproduced by {\bf the} NLFFF model with a weakly twisted flux rope at the core. 
	
\tiny
 \keyFont{ \section{Keywords:} 
 	Sun:flares, Sun:coronal mass ejection, Sun: magnetic fields, Sun: magnetic reconnection, Sun: sigmoid, } 
\end{abstract}

\section{Introduction}
\label{Intro}

Most often the coronal mass ejections are seen to launch from magnetically concentrated regions called active regions (ARs).  In soft X-rays or in EUV images, the ARs that precede CMEs are seen with the observational features of S- and J-shaped loops situated over along the polarity inversion line (PIL). Owing to this specific S- or inverted S-shape, \citet{rust1996} termed these regions as sigmoids and are considered to be one of the most important precursor structures for the solar eruptions \citep{hudson1998,canfield1999,canfield2007}.  

The shape of the sigmoid indicates the loops composed of non-potential magnetic field configuration characterised by sheared and/or twisted magnetic field lines. As a reason, the magnetic structure of the sigmoids are described by two competing configurations that are sheared arcade and magnetic flux rope (MFR). In the sheared arcade model, the two magnetic elbows sheared past each other are situated at the opposite ends of the PIL with a central sheared core at the middle section of the PIL \citep{moore2001,moore2006}. And in the MFR scenario, a magnetic MFR is embedded in a stabilizing potential envelope field \citep{moore1992, hood1979, rust1996}. Because of the non-potential nature of the magnetic field, the magnetic structure in the sigmoid is approximated with a non-linear force-free field (NLFFF) which allows different twist parameter for individual field lines. The force-free modeling is justified by the low-$\beta$ corona with a slowly evolving field compared to the Alfven crossing time. 

\begin{figure}[!ht]
	\centering
	\includegraphics[width=0.99\textwidth]{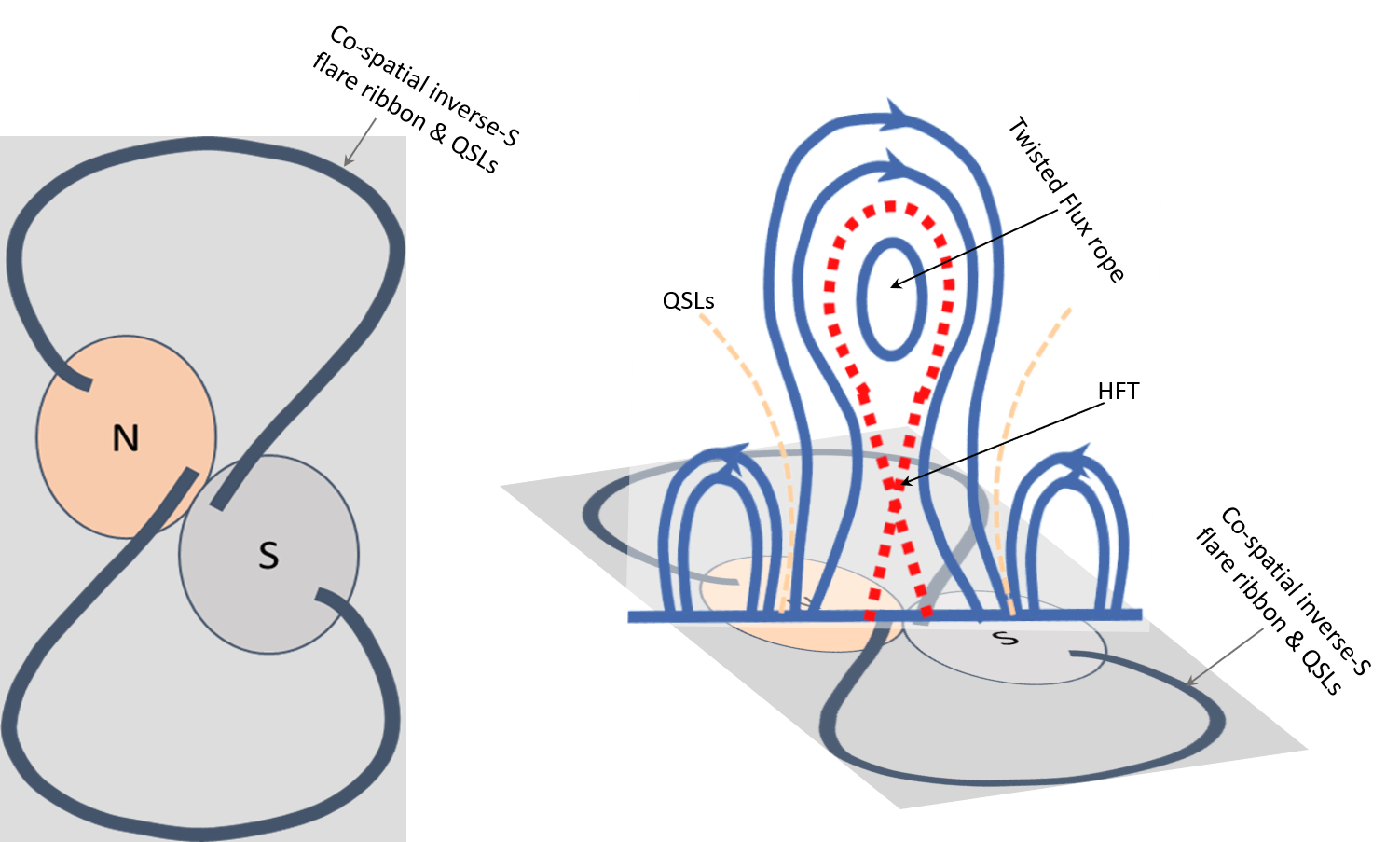}
	\caption{ Schematic of the 3D standard model for eruptive flares. {\bf left panel:} Black thick curves indicate the inverse-S QSL footprints at the photosphere. ``N" (``S") refers to north (south) polarity magnetic flux distribution. These inverse-S QSLs are regarded as the combination of two inverse-J shaped sections with straight parts lying in the opposite polarities about the PIL.  {\bf right panel:} Perspective view of the erupting flux rope structure and HFT underneath in a vertical plane across the MFR. Erupting MFRs form flare ribbons which trace these photospheric QSL footpoints. }
	\label{fig_cartoon}
\end{figure}

\begin{figure}[h!]
	\centering
	\includegraphics[width=\textwidth]{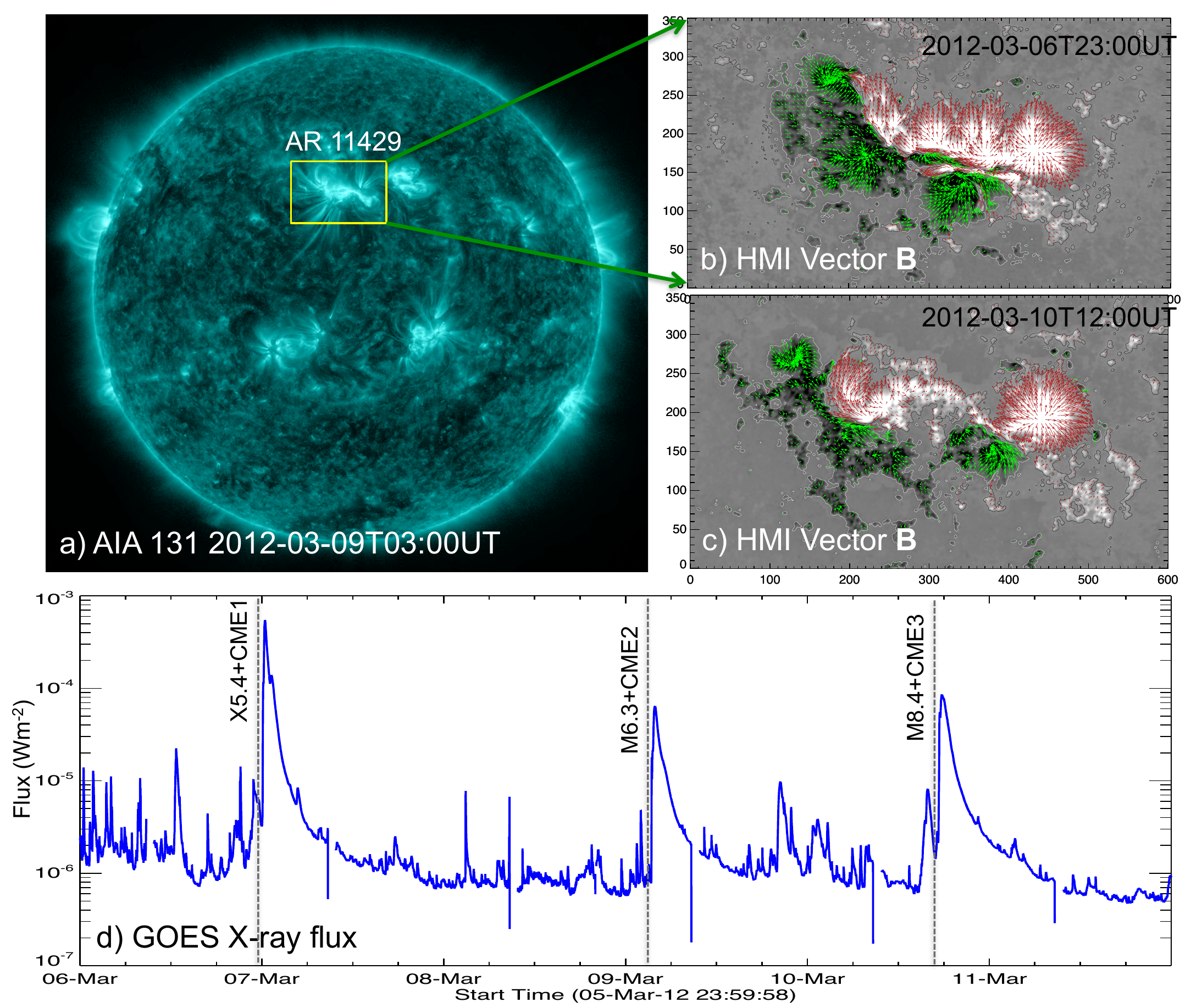}
	\caption{Observations of the successively erupting AR 11429 {\bf a)} Full disk observation of AIA 131~\AA. The structured emission from AR 11429 is outlined with yellow rectangular box, {\bf b-c)} HMI vector magnetogram observations of the AR 11429 at the onset of CME eruptions on March 6 and 10 respectively. The background image is vertical magnetic field with contours at $\pm$120 G. Green/red arrows denote horizontal field with their length being proportional to the magnitude $B_h=\sqrt(B_x^2+B_y^2)$. Axis units are pixels of 0.5 arc-second, {\bf d)} Disk integrated GOES X-ray flux in 1-8~\AA~band during the disk transit time of the AR 11429. Text near the X-rays peaks refers to the CMEs and associated X, M-class flares from the AR 11429. }
	\label{fig_11429}
\end{figure}

To understand the eruptive nature of sigmoidal regions, it is important to study their magnetic structure and evolution both theoretically and observationally. Observations showed that the magnetic flux ropes have frequently been associated with sigmoidal regions \citep{rust1996,gibson2006,Green2007,SongHQ2015, vemareddy2015_FullStudy_SunEarth, SongHQ2018} and the modelled NLFFF magnetic configuration that describes the observed shape of sigmoids is a weakly twisted flux rope embedded in potential arcade \citep{savcheva2012a,zhaoj2016, Guoy2013,Vemareddy2016_sunspot_rot, vemareddy2018}. The flux rope structure constructed from analytical configurations \citep{Demoulin2006_qsl_adsr,Demoulin2007_cursheet_adsr} exhibits current sheets in the magnetic interface layers called quasi-separatrix layers (QSLs) where the connectivity of the field lines changes drastically just like separatrix layers. In the process of emerging, such a MFR develops a separatrix surface touching the photosphere along the PIL section \citep{titov1999, titov2002}. These sections of the PIL are called bald-patches (BPs) and the separatrix surface at the BP appears as an S-shape from top view similar to sigmoid shape. After the emergence phase, the S-shaped bald-patch separatrix surface (BPSS) bifurcates into a double J-shaped QSL with the main body of the MFR lifted off. From this BPSS topology, the QSL structure underneath the rising flux rope develops an X-line configuration referred to as hyperbolic flux tube (HFT), where the reconnection sets in for the onset of the eruption. Therefore, the HFT topology is the predicted site for flare reconnection and CME eruption \citep{savcheva2012c}. Further, the topological analysis of the MFR configurations, both models and observations, recommends the extension of the standard CHSHK flare model \citep{carmichael1964, sturrock1966,Hirayama1974,KoppPneuman1976} to 3D. In the 2D flare model two flare ribbons are observed on either side of the PIL, whereas MFR eruptions found co-spatial flare ribbons with hook-shaped QSLs \citep{Pariat2012,Janvier2013,janvier2015,zhaoj2016}.  Figure~\ref{fig_cartoon} displays the schematic of the 3D standard model for eruptive flares as interpreted from simulations of eruptive MFRs \citep{titov1999,Pariat2012,ChenPF2012_WheredoFlareRibbons_Stop}. In the left panel, the hook-shape inverse-S shaped QSLs in the photosphere are shown in thick black curves. The HFT underneath the uplifting MFR is depicted in the right panel as seen from perspective view. The photospheric QSL footpoints are co-spatial with flare ribbons in the 3D-eruptive flare models which is a signature of the MFR topology. 

Motivated by the above topological studies of the erupting regions, in this paper, we study the pre-eruptive magnetic structure of five CMEs from two successively erupting ARs. The coronal field is constructed by NLFFF \citep{wiegelmann2004}, then we computed the chromospheric QSLs to compare their spatial locations with the geometry of the observed flare ribbons. By this comparison, one can ascertain the model predictions of the topological features with the observations, and then also validate the extent of the invoked NLFFF model to reproduce the actual coronal field. The paper is organized as follows. In section~\ref{sec_OverView}, an overview of the CME events with a brief description of results presented in the previous reports. Reconstruction of the magnetic structure by NLFFF modeling is presented in Section~\ref{sec_ModMagStr}. Comparison of the QSLs with the flare-ribbons is made in Section~\ref{sec_qsl_comp} and concluded with a summarized discussion in Section~\ref{sec_Summ}.

\begin{figure}[h!]
	\centering
	\includegraphics[width=\textwidth]{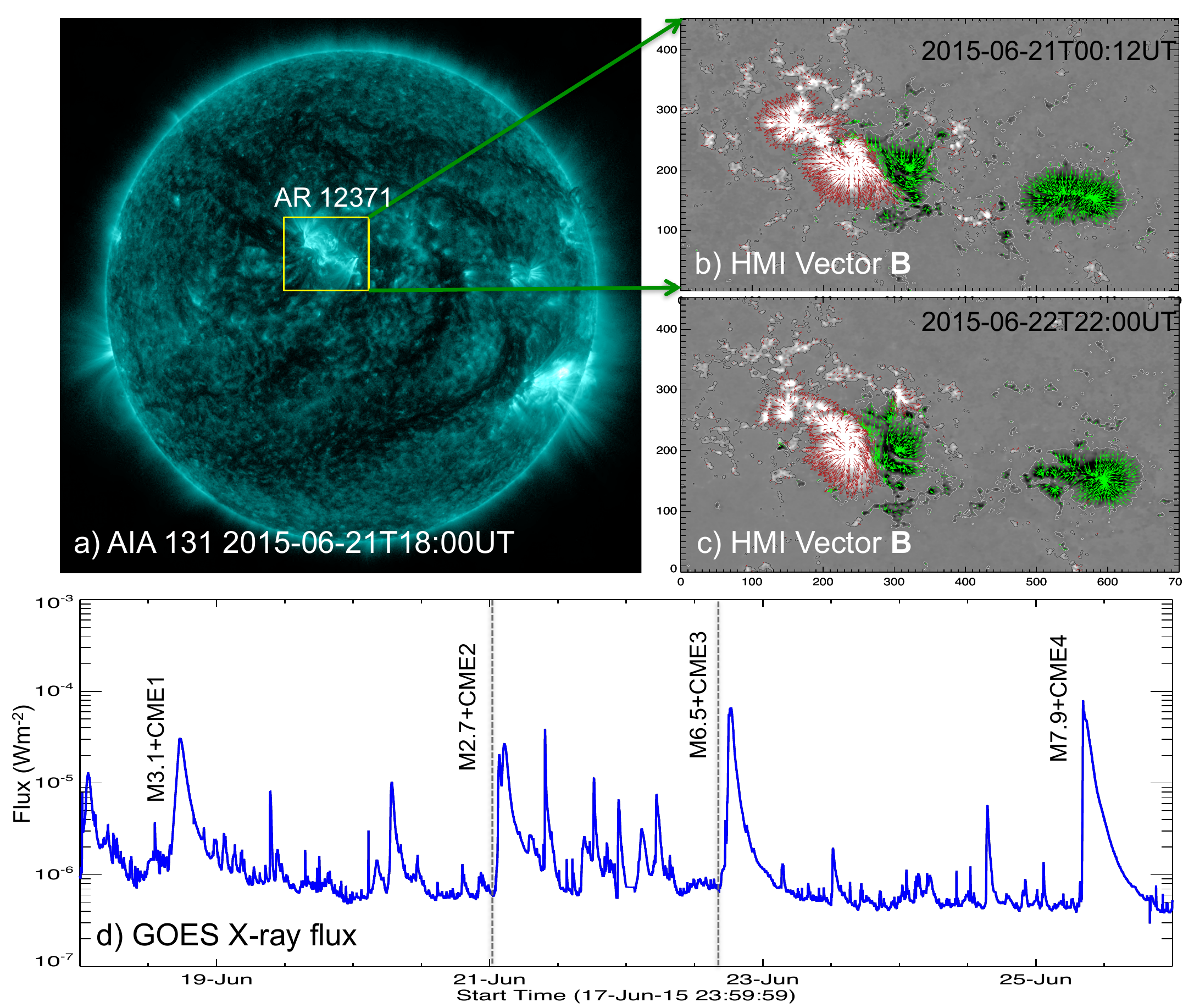}
	\caption{Observations of the successively erupting AR 12371 {\bf a)} Full disk observation of AIA 131~\AA. The structured emission from AR 12371 is outlined with yellow rectangular box, {\bf b-c)} HMI vector magnetogram observations of the AR 12371 at the onset of CME eruptions on June 21 and 22 respectively. The background image is vertical magnetic field with contours at $\pm$120 G. Green/red arrows denote horizontal field vector whose length is proportional to magnitude $B_h=\sqrt(B_x^2+B_y^2)$. Axis units are pixels of 0.5 arc-second units, {\bf d)} Disk integrated GOES X-ray flux in 1-8~\AA~band during the disk transit time of the AR 12371. Text near the X-ray peaks refers to the CMEs and associated M-class flares from the AR 12371. The events marked with vertical lines are studied in this work. }\label{fig_12371}
\end{figure}

\section{Overview of the Observations}
\label{sec_OverView}
Observations of the successive eruptions from the ARs 11429 and 12371 are obtained from the Atmospheric Imaging Assembly (AIA; \citealt{lemen2012}) and Helioseismic Magnetic Imager (HMI; \citealt{Schou2012}) on board NASA's space-based Solar Dynamics Observatory (SDO). The AIA instrument captures the full disc images of the solar corona in 10 wavelengths at 0.6 arcsec pixel$^{-1}$ resolution. The HMI provides photospheric line of sight and vector magnetic field observations at 0.5 arcsec pixel$^{-1}$ resolution in Fe {\sc i} 6173\,\AA~wavelength. The pipeline procedures of deriving the vector magnetic fields from the Stokes filter images are documented in \citet{bobra2014} and \citet{hoeksema2014}. We used vector magnetic field data product \texttt{hmi.sharp\_cea\_720s} at a cadence of 720s. Additional information of the CME eruptions is obtained from web portals like CME catalog\footnote{\url{https://cdaw.gsfc.nasa.gov/CME_list/}}, and solar monitor\footnote{\url{https://www.solarmonitor.org}}.   

The AR 11429 was a pre-emerged one probably on far-side of the sun.  Figure~\ref{fig_11429} demonstrates the eruption scenario in this AR 11429. In panel~\ref{fig_11429}a. the full disk image of AIA 131~\AA~shows the structured coronal emission from the AR 11429 (yellow rectangular box) located in the north (N17$^\circ$) hemisphere. HMI observations of the photospheric vector magnetograms at the pre-eruption time on two different days are displayed in panels~\ref{fig_11429}(b-c). The AR presents a large interface opposite polarities known as polarity inversion line (PIL). These magnetic polarities evolve with persistent shearing and converging motions which led to continuous flux cancellation as observed with the decay of the magnetic flux regions. Such regions are favorable to form stored energy configurations of sheared magnetic fields along the PIL. As a result, the AR evolved to a much more complex magnetic configuration  ($\beta/\gamma/\delta$) producing  severe flare/CME activity during its disk transit. In Panel~\ref{fig_11429}(c), GOES X-ray flux (1-8~\AA~band) displays the peaks of one X and two M-class flares associated with the CME eruptions from the AR 11429. Because of recurrent eruptions from the same region under a continuous physical process, these eruptions are referred to as homologous eruptions \citep{nitta2001,ZhangJun2002_HomoCME}. Study of this AR by \citet{DhakalS2020} suggests that the shearing motion and magnetic flux cancellation by converging fluxes were key processes to recurrently form the erupting structure and then its eruption. Details of the CME eruptions from this AR are listed in Table~\ref{tab1}.

Another recurrent CME producing AR was NOAA 12371, which was also a pre-emerged region that passes the visible solar disk 12$^\circ$N on June 16, 2015. Figure~\ref{fig_12371} presents the eruption scenario of this AR. A representative AIA 131~\AA~full disk image is displayed in panel~\ref{fig_12371}a, which shows the sigmoidal morphology of AR corona (yellow rectangular box). Panels~\ref{fig_12371}(b-c) display the HMI vector magnetograms at the pre-eruption time on June 21 and 22. The AR essentially consists of a leading negative flux region with the following interacting opposite polarity regions. From these magnetic field observations, the AR's disk passage reveals that the following bipolar region was seen with large shear and converging motion as a result the flux distribution becomes diffused and disintegrated in successive days. As in the earlier AR, such an evolution of magnetic polarities leads to formation of the twisted flux along the PIL, which indeed is revealed by the sigmoidal loop structure in the EUV images. Four major CME eruptions occurred associated with M-class flares from this AR as shown by the peaks in the GOES X-ray flux plot of panel~\ref{fig_12371}c. A detailed study of this AR by \citet{Vemareddy2017} interprets the successive eruptions by the cyclic process of energy and helicity storage over a time scale of a day or two and then its release by the CME eruptions. Table~\ref{tab1} lists the CME eruptions and associated flares from this AR. 

\begin{table}[]
	\centering
	\caption{Information of CME eruptions and associated flares from the studied ARs}
	\begin{tabular}{lllll}
		\hline\hline    
		Event & Flare peak time [UT] & Location & Associated Flare [UT] & CME speed [Km/s]  \\
		\hline\hline 
		&       &   \textbf{AR NOAA 11429}   &       &  \\
		\hline
		CME1 & SOL2012-03-07T00:24 & N17$^\circ$E31$^\circ$ & X5.4 (00:24 ) & 1825 \\
		{CME2} & SOL2012-03-09T03:53 & N15$^\circ$W03$^\circ$ & M6.3 (03:53) & 950 \\
		CME3 & SOL2012-03-10T17:44 & N17$^\circ$W24$^\circ$  & M8.4 (17:44) & 1296 \\
		\hline
		&        &   \textbf{AR NOAA 12371}    &       &  \\
		\hline     
		CME1 & SOL2015-06-18T17:35  &  N10$^\circ$E50$^\circ$  & M3.1(17:35) &  1305        \\
		CME2 & SOL2015-06-21T02:36  & N12$^\circ$E16$^\circ$      & M2.2 M2.7 (02:36) & 1366 \\
		CME3 & SOL2015-06-22T18:23    & N13$^\circ$W06$^\circ$      & M6.5 (18:23) & 1209 \\
		CME4 & SOL2015-06-25T08:16    & N12$^\circ$W40$^\circ$      & M7.9 (08:16) & 1627 \\
		\hline\hline
	\end{tabular}
	\label{tab1}
\end{table}%

\section{Modelled magnetic structure: non-Linear force-free field}
\label{sec_ModMagStr}

The AR magnetic structure in 3D is modelled by applying the NLFFF extrapolation \citep{Wheatland2000,wiegelmann2004,wiegelmann2010}. The NLFFF algorithm involves minimization of the functional 

\begin{eqnarray} L &=&\int _{V}{\left(w\frac{|(\nabla \times \mathbf {B})\times \mathbf {B}{{|}^{2}}}{{{B}^{2}}}+w|\nabla \bullet \mathbf {B}{{|}^{2}} \right)}dV\nonumber\\ &&+ \nu \int _{S}{\left(\mathbf {B}-{{\mathbf {B}}_{{\rm obs}}} \right)}\cdot W \cdot \left(\mathbf {B}-{{\mathbf {B}}_{{\rm obs}}} \right)dS. 
\label{eq_nlfff}
\end{eqnarray}

In the above equation, the first integral includes quadratic forms of the force-free and solenoidal conditions and $w$ is a weighting function toward the lateral and top boundaries. Second term is surface integral to take into account measurement errors and allows a slow injection of the boundary data controlled by the Lagrangian multiplier $\nu$ (see, \citealt{wiegelmann2010} for more details). $W(x, y)$ is a diagonal matrix, which is chosen inversely proportional to the transverse magnetic field strength. 

\begin{figure}[h!]
	\centering
	\includegraphics[width=0.9\textwidth]{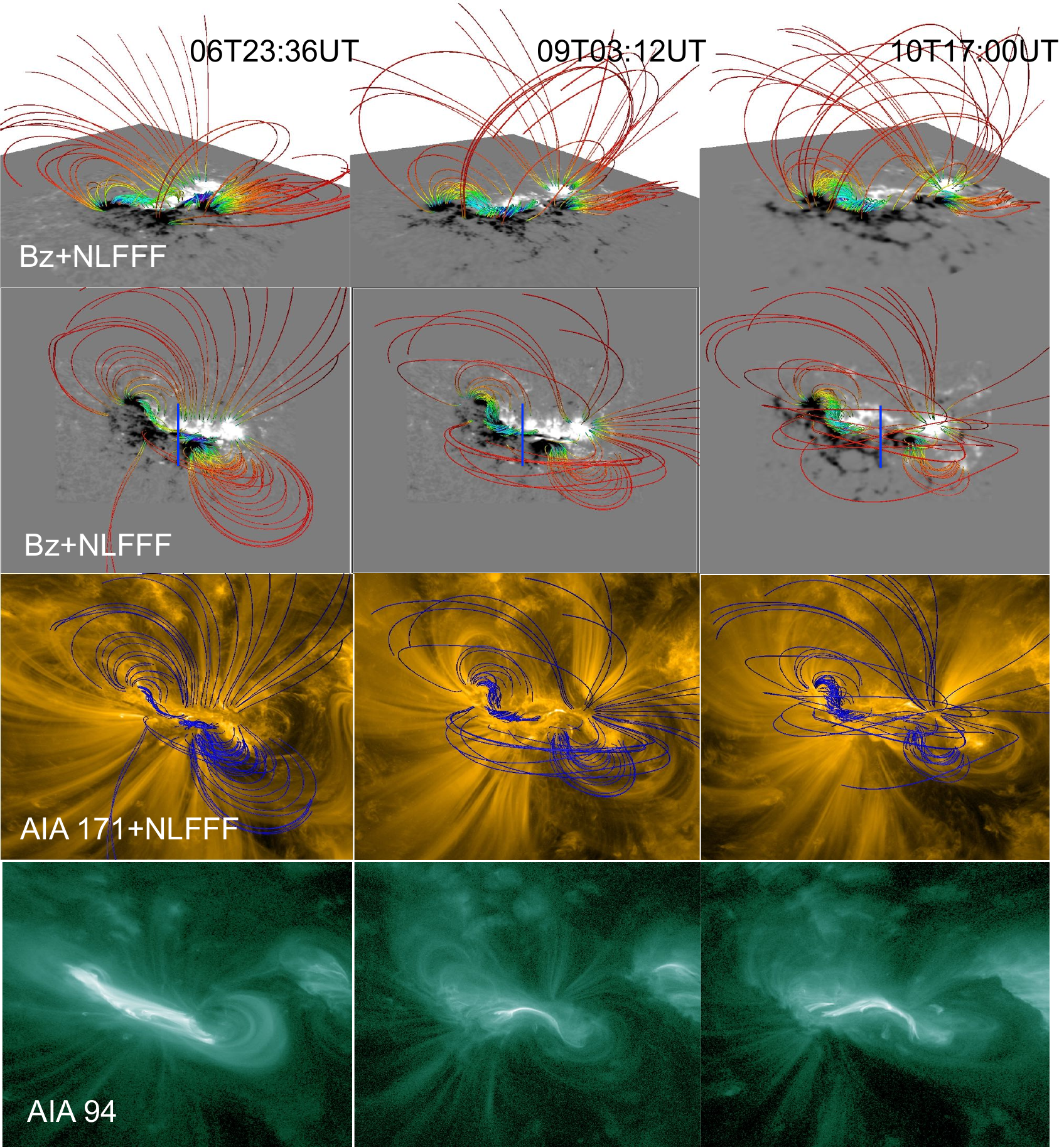}
	\caption{Pre-eruptive magnetic structure modeled by NLFFF in AR 11429 for the three CME events. {\bf First row:} Perspective view of the field-line rendering. Bottom image is photospheric magnetogram ($B_z$). Twisted structure along the PIL is overlaid by the less sheared field lines {\bf Second row:} field lines plotted on top of the photospheric magnetogram (Bz). The global structure is a sigmoid with two opposite-J-shaped loops and an inner core of highly twisted field. Vertical blue lines indicate the location of the slice plane for QSL analysis.  {\bf third row:} same field lines on AIA 171~\AA~images, {\bf fourth row:} AIA 94~\AA~ images at the same time. Hot plasma emission is mainly co-spatial with the strongly sheared core along the PIL. To a good approximation, the modeled structure resembles the coronal plasma loops in the EUV images. Field lines are color coded (blue (red): ~1200 (2)G) with the horizontal field strength in height in first and second rows.}\label{fig_ext_11429}
\end{figure}

The observed photospheric vector magnetic field from the HMI is used as the lower boundary condition, for which the field has to satisfy flux balance, force-free conditions. When the vector magnetic field has full coverage of the AR, the flux balance condition is normally satisfied within 10\% deviation. To keep this value as small as possible, we multiply the positive polarity region with a relative flux factor defined by the ratio of positive and negative flux. The force-free conditions are further satisfied by applying a pre-processing procedure on the magnetic field components \citep{wiegelmann2006}. To facilitate tracing field lines in a large extent of volume comparable to EUV field-of-view, the boundary magnetic field observations are inserted in an extended field of view and then rebinned to 1 arcsec pixel$^{-1}$. With the normal field component of magnetic field, we reconstructed the 3D potential field (PF) which is then used as the initial condition and also to prescribe the top and side boundaries for the NLFFF algorithm. 

\begin{figure}[h!]
	\centering
	\includegraphics[width=11.5cm]{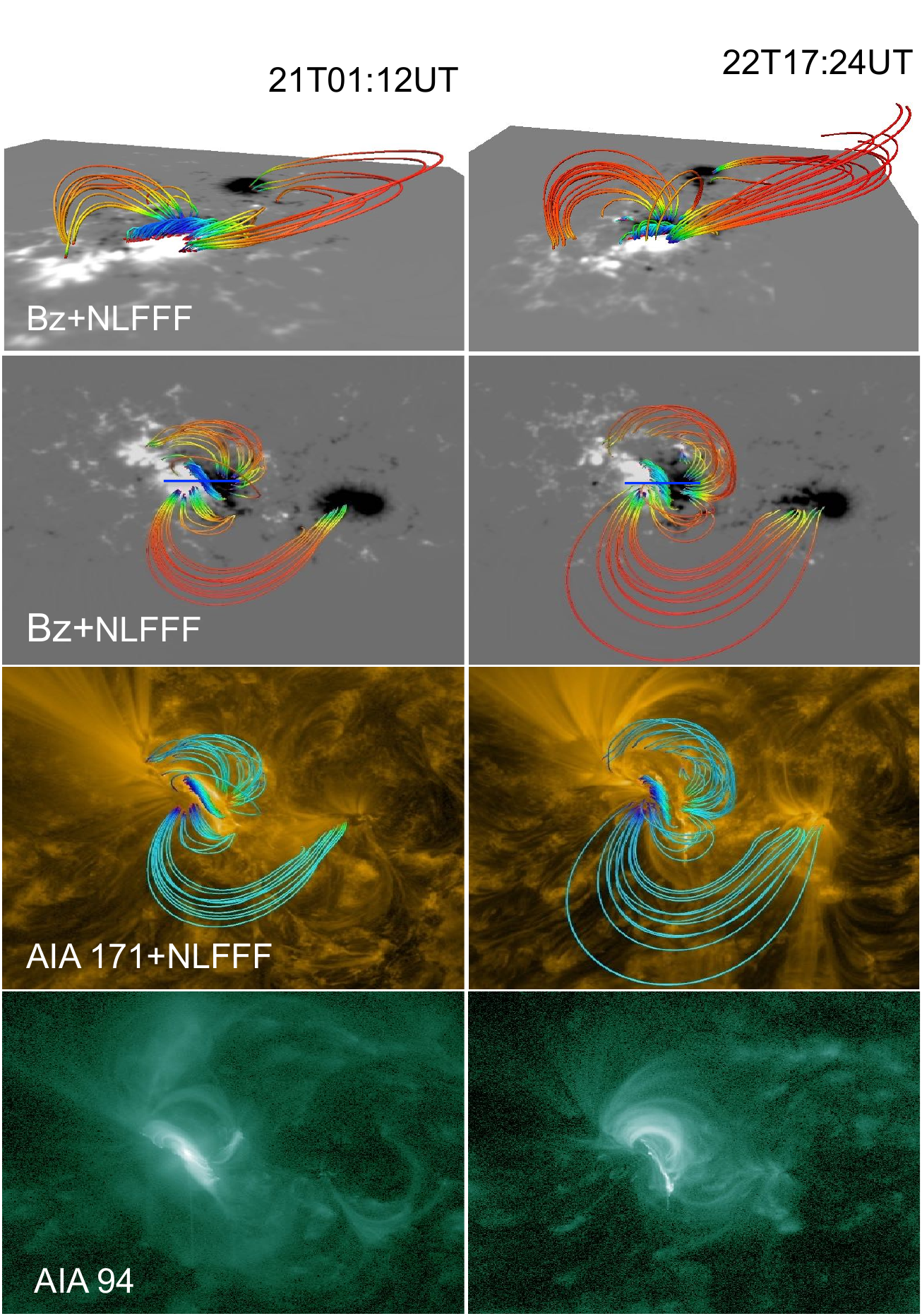}
	\caption{Pre-eruptive magnetic structure modeled by NLFFF in AR 12371 for the CME1 and CME2 events. {\bf First row:} Perspective view of the field-line rendering. Bottom image is photospheric magnetogram ($B_z$). Twisted structure along the PIL is overlaid by the less sheared field lines {\bf Second row:} field lines plotted on top of the photospheric magnetogram ($B_z$). The global structure is a sigmoid with two opposite-J-shaped sections and an inner core of highly twisted field. Horizontal blue lines indicate the location of the slice plane for QSL analysis. {\bf third row:} same field lines on AIA 171~\AA~images, {\bf fourth row:} AIA 94~\AA~ images at the same time. Hot plasma emission is mainly co-spatial with the strongly sheared core along the PIL. To a good approximation, the modeled structure resembles the coronal plasma loops in the EUV images. Field lines are color coded (blue (red): ~1200 (2)G) with the horizontal field strength in height in first and second rows.}\label{fig_ext_12371}
\end{figure}

In Figure~\ref{fig_ext_11429}, we show the NLFFF magnetic structure of AR 11429 just before the CME events on March 6, 9 and 10. The NLFFF is constructed on a grid of of $450\times450\times201$ representing the physical dimensions of $328\times328\times146$ Mm$^3$ AR corona. Similarly in Figures~\ref{fig_ext_12371}, we show the NLFFF magnetic structure of AR 12371 just before the CME events on June 21 and 22. In this case, the NLFFF field is constructed on a grid of of $512\times512\times256$ representing the physical dimensions of $373\times373\times186$ Mm$^3$ AR corona. For all of the cases, the NLFFF relaxation converges to an average field divergence of the order $10^{-4}$ and an average field-aligned current defined by $\theta_J$ to an extent 9-12$^\circ$. 

To capture the most sheared structure, the field lines are rendered according to total current density ($|\mathbf{J}|$) and horizontal field component ($B_h$) at the bottom boundary. The field line rendering is shown in  perspective and top views in 1st and 2nd rows respectively of Figures~\ref{fig_ext_11429} and~\ref{fig_ext_12371}. In these panels, the bottom plane is an observed $B_z$ map. The field rendering in these panels comprise two inverse J-shaped field lines sheared past each other about the main PIL, which together reveal the shape of the inverse S-sigmoid. Owing to shearing motions of the foot points, the field lines near PIL are strongly stressed, manifesting low lying twisted core of the sigmoid, which is regarded as flux rope with helical field lines. During the onset of the eruption, this flux rope builds up further by the reconnection of oppositely sheared field, and therefore is the central structure of the solar eruptions \citep{moore2001, DuanAiying2019_MFRs}.

In order to judge the NLFFF model to the coronal magnetic field, the modeled magnetic structure is compared to EUV coronal observations. To this end, we ensure that the EUV images are co-aligned to the magnetic field with the same field-of-view. The same field line rendering is over-plotted on EUV observations of corona captured in AIA 171~\AA~images(second row panels) which delineates a good global resemblance of the field lines with the plasma tracers. Because of the  strong volume currents, the intense hot emission in AIA 94~\AA~images is mostly co-spatial with the NLFFF twisted core along the PIL.

\begin{figure}
    \centering
    \includegraphics[width=1.0\textwidth]{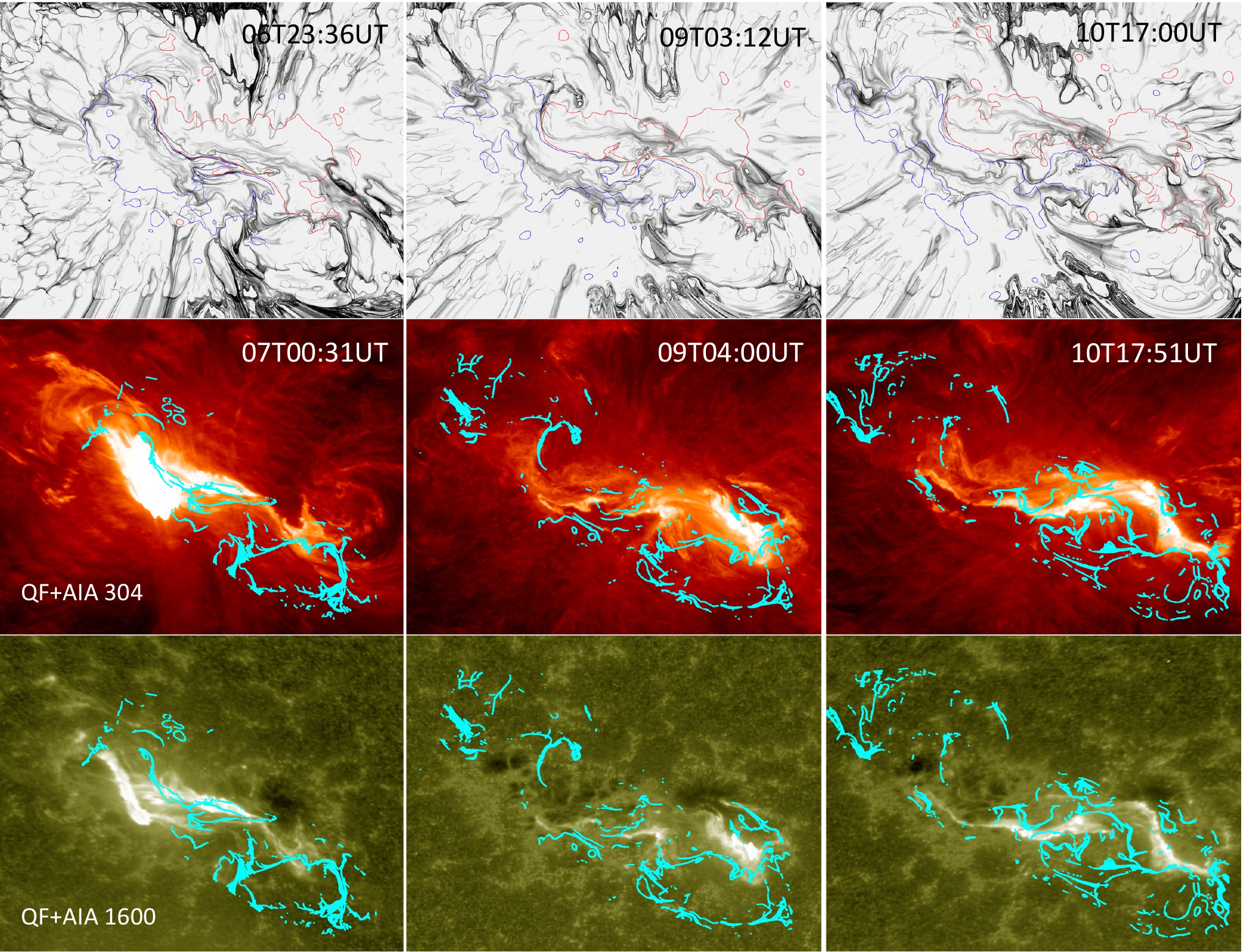}
    \caption{Comparison of QSLs and flare ribbons for the pre-eruptive magnetic structure of CME1 (first column), CME2 (second column), CME3 (third column) in the AR 11429. \textit{First row:} Inverse maps of $Log (Q)$ obtained at $z=1.5 Mm$. $B_z$-contours at $\pm 120$ G are overdrawn (red/blue curves). QSLs with large Q values are identified by intense black traces in strong field region. All maps are scaled within $1<Log(Q)<7$. QSLs of large Q-values separate the sheared/twisted core field from the surrounding less sheared field in the following bipolar region. Also higher Q-values in quiet regions are due to noisy transverse field and have no relevance to the magnetic structure of interest. {\bf Second and third rows:} Co-spatiality of QSLs and flare ribbons. Contours of $Log (Q)=[4, 5, 6]$ (in cyan color) on AIA 304~\AA~(second row) and AIA 1600~\AA ~(third row) snapshots taken at around the peak flare time. QSLs are co-spatial with the flare ribbons including the hooked shape.}
    \label{fig_qf_11429}
\end{figure}

\begin{figure}[h!]
	\centering
	\includegraphics[width=0.7\textwidth]{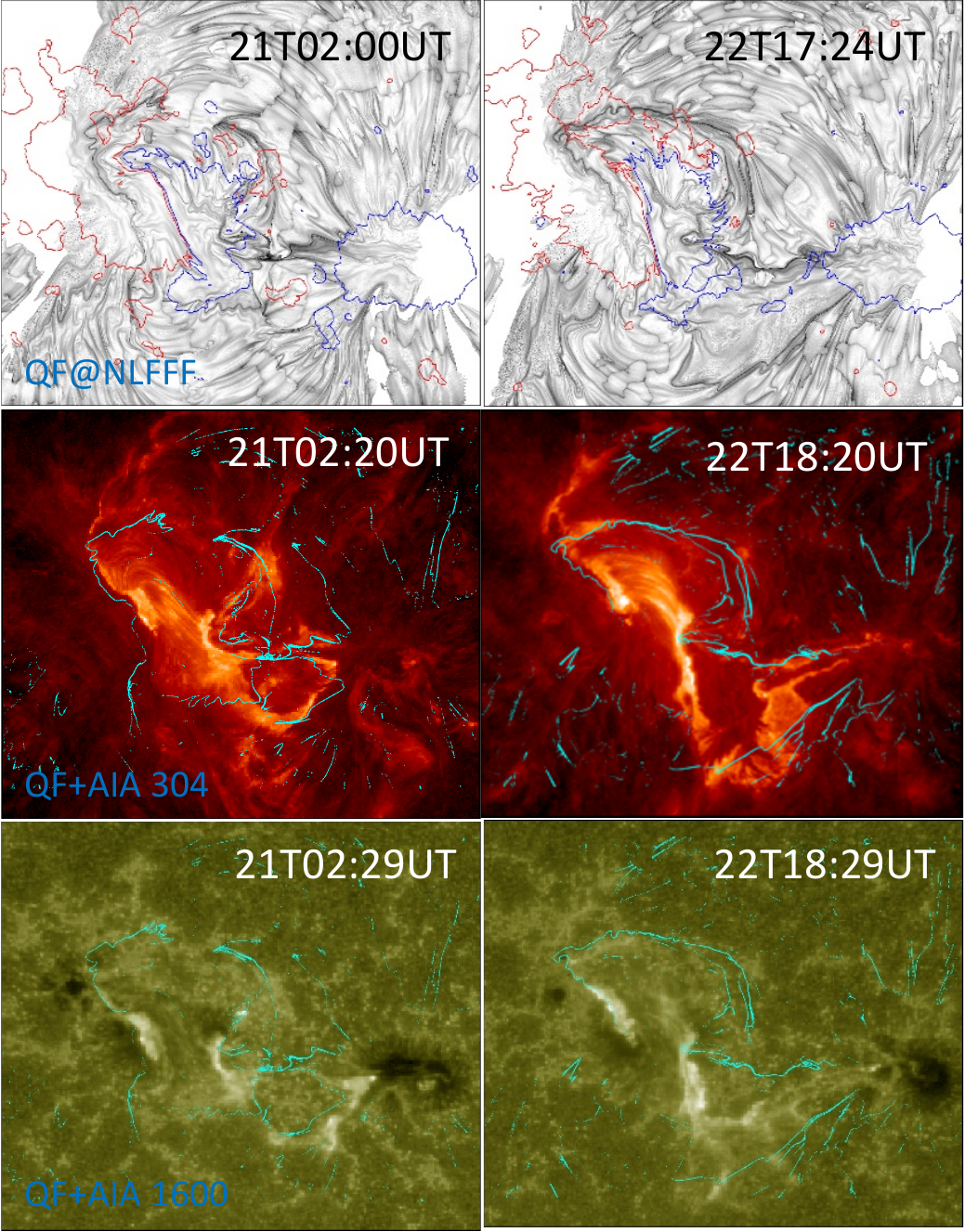}
	\caption{Comparison of QSLs and flare ribbons for the pre-eruptive magnetic structure of CME1 (first column) and CME2 (second column) in the AR 12371. \textit{First row:} Inverse maps of $Log (Q)$ obtained at $z=1.5 Mm$. $B_z$-contours at $\pm 120$ G are overdrawn (red/blue curves). QSLs with large Q values are identified by intense black traces in strong field region. All maps are scaled within $1<Log(Q)<7$. QSLs of large Q-values separate the sheared/twisted core field from the surrounding less sheared field in the following bipolar region. Also higher Q-values in quiet regions are due to noisy transverse field and have no relevance to the magnetic structure of interest. {\bf Second and third rows:} Relation between QSLs and flare ribbons. Contours of $Log (Q)=[5, 6]$ (in cyan color) on AIA 304~\AA~ (second row) and AIA 1600~\AA~(third row) snapshots taken at around the peak flare time. The contours constitute two inverse-J sections, qualitatively outlining the inverse-S flare ribbons. }\label{fig_qf_12371}
\end{figure}

\section{Quasi-separatrix layers and flare ribbons}
\label{sec_qsl_comp}
QSLs are the regions of the magnetic volume where the field line connectivity experiences dramatic but continuous variations, including possible discontinuities in the mapping, so are the generalized features to the separatrices \citep{demoulin1996}. From the constructed 3D coronal fields, the change in magnetic field line linkage in the volume is measured by the strength of QSLs which is defined by squashing factor Q \citep{titov2002,titov2007}. Q describes the gradients in the field line mapping whose larger values correspond to the cross section of QSLs in any plane. It is computed by tracing two consecutive field lines with foot points at an extremely small distance and then measuring the distance between the respective conjugate foot points as given by the following mathematical expression

\begin{equation} 
Q=\frac{{{\sum _{i,j=1}^{2}{\left(\frac{\partial {{X}_{i}}}{\partial {{x}_{j}}} \right)}}^{2}}}{|{{B}_{z,0}}/{{B}_{z,1}}|}, 
\label{eq_qf}
\end{equation}

where $X_i$(i = 1, 2) is the coordinates of the conjugate foot point in the Cartesian system and $B_{z, 0}$ and $B_{z, 1}$ are vertical field components at the starting and ending footpoints of a field line. From the 3D NLFFF, we calculate Q using the code developed by \citet{LiuRui2016} according to formalism prescribed in \citet{Pariat2012}. The field lines are traced by integrating the first order differential equations by 4th order Runge-Kutta solver with the help of tri-linear interpolation scheme. To have a smooth and dense distribution of Q, these computations are performed on a finer grid of resolution increased by eight times that of the extrapolation grid. 

The complexity in the QSL maps of the NLFFF model is intrinsic to the large amount of fragmentation in the observed photospheric magnetic-field distribution. It has been shown that the complexity of QSL maps decreases with the height \citep{savcheva2012a, savcheva2012c}. This is a consequence of the magnetic field being progressively smoother with height. Therefore, to determine important QSLs, Q maps are computed at 1.5 Mm height above the photosphere, which we refer chromospheric QSLs in this manuscript discussion. 

In the first row panels of Figures~\ref{fig_qf_11429} and \ref{fig_qf_12371}, we plot $\log Q$ maps at $z=1.5\,Mm$ with inverse sign, so the darker parts correspond to higher $Q$ values. For a reference, $B_z$ contours of $\pm 120$G are over-plotted. Due to noisy transverse field, the Q maps of NLFFF are more fragmented in weak field regions. The relevant QSLs are having large Q values ($>10^7$) located in the stronger magnetic polarities related to the sigmoid structure. In panel of 23:36UT on March 6, the the relevant QSLs traces the PIL in addition to the QSLs in the hooked sections in the top and bottom of the sigmoid. A similar distribution is presented in other panels. As seen in the 02:00 UT map on June 21, the QSL of high-Q values have two sections each falling in the opposite polarity regions of inner bipolar regions, both of which extends towards leading negative polarity. These two sections joined at the top in positive polarity. Although lower part have smaller values of Q, this overall QSL is continuous and encloses an inverse S-shaped region filled with lower Q-values. Essentially, these QSL highlights the difference between the twisted core flux within the region and the surrounding potential like arcade outside. 

\begin{figure}[!ht]
    \centering
    \includegraphics[width=0.99\textwidth]{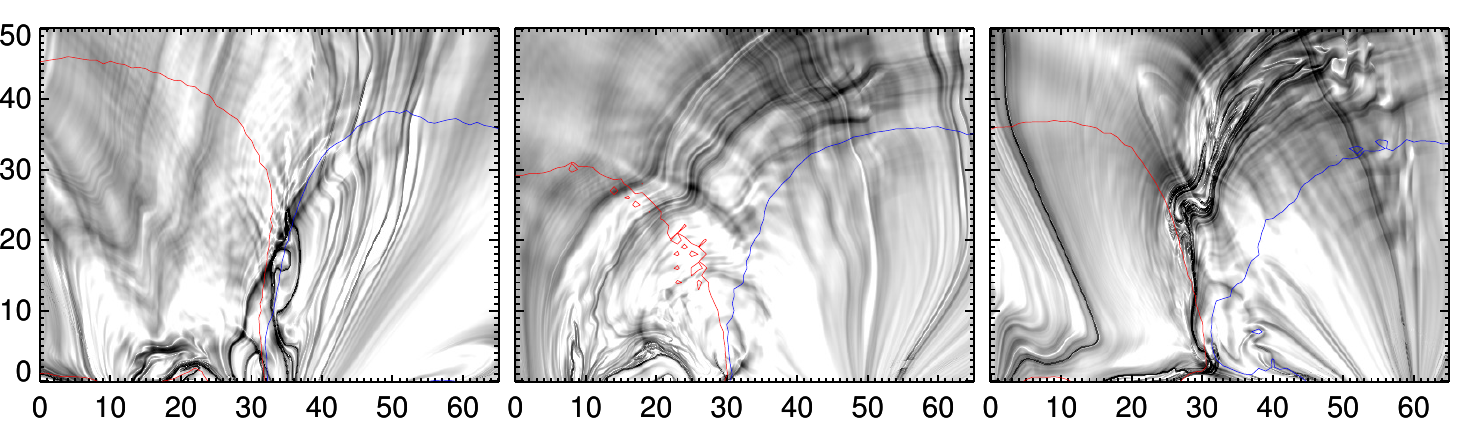} 
    \caption{Inverse maps of $Log(Q)$ in the vertical slice placed across the sigmoid in the AR 11429 as shown in Figure~\ref{fig_ext_11429}. Contours of $B_z$ ($\pm$80\,G, red/blue curves) are over plotted. QSLs of large Q-values (black) are boundaries of magnetic domains enclosing the twisted-sheared core about the PIL.}
    \label{fig_qf_vert_11429}
\end{figure}

\begin{figure}[!ht]
    \centering
    \includegraphics[width=0.99\textwidth]{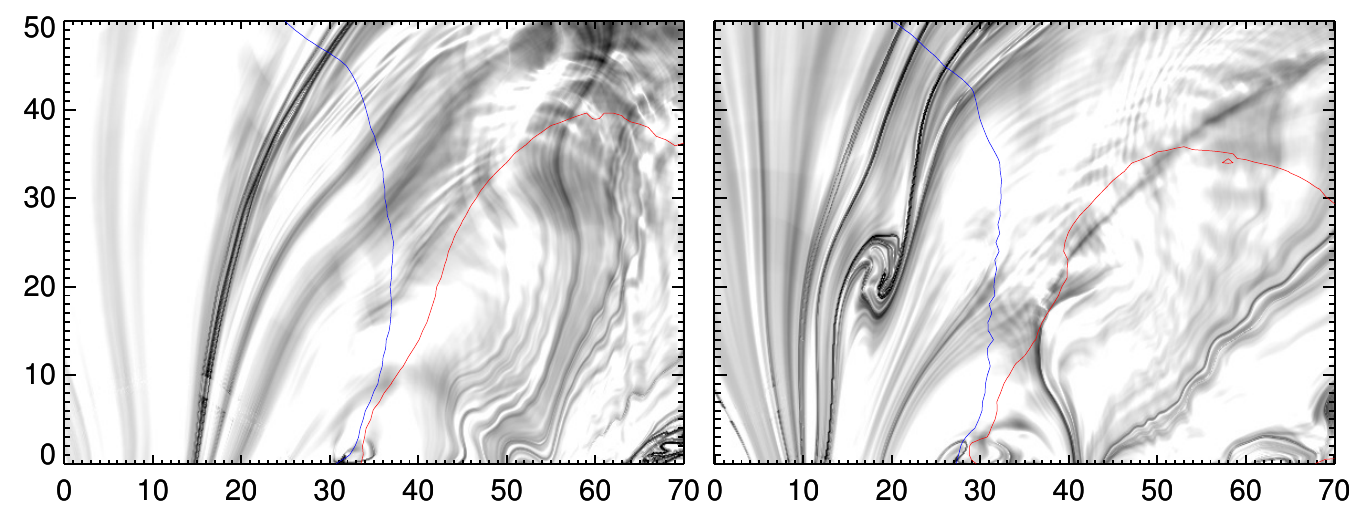}
    \caption{Inverse maps of $Log(Q)$ in the vertical slice placed across the sigmoid in the AR 12371 as shown in Figure~\ref{fig_ext_12371}. Contours of $B_z$ ($\pm$80\,G, red/blue curves) are over plotted. QSLs of large Q-values (black) are boundaries of magnetic domains enclosing the twisted-flux core about the PIL. (mark the MFR/twisted flux with an arrow) }
    \label{fig_qf_vert_12371}
\end{figure}

This picture is even clearer in the vertical cross-section map of Q obtained across the sigmoid. These are shown in Figures~\ref{fig_qf_vert_11429} and~\ref{fig_qf_vert_12371}. $B_z$-contours ($\pm 80$G, red for positive, blue for negative), obtained in the same cut-plane, are over-plotted. Different from horizontal maps, the vertical Q maps are smooth with an obvious relevance with the magnetic polarities about the PIL. In these maps, the QSLs of large Q values well distinguish two closed domains belonging to largely sheared sections on either side of the PIL and the surrounding potential like arcade. Importantly the QSLs intersect the magnetic polarities at the middle in the NLFFF, whereas they cover the entire polarity region in the PF. This is clearly an indication of a sheared core being surrounded by less-sheared/potential arcade in the AR. Given the twisted flux (flux rope) at the core, the QSLs in its cross section mimics a inverse tear-drop shape as predicted by the theoretical models quoted in the Introduction. Depending on the degree of the twist, the identification of such QSLs varies due to observational and modeling difficulties (see for example Figure 8 in \citealt{vemareddy2019_VeryFast} and Figure 4 in  \citealt{zhaoj2016}). In our NLFFF structures of the five CMEs, it is mildly visible as shown in Figure~\ref{fig_qf_vert_11429} and \ref{fig_qf_vert_12371}. 


For a comparison with flare ribbons, the corresponding AIA 304~\AA\, 1600~\AA observations are displayed in the second and third row panels of Figures~\ref{fig_qf_11429} and \ref{fig_qf_12371} respectively. On these maps, contours of Q for the corresponding event at $10^5$, $10^6$ levels are over-plotted in cyan-color. Note that we are comparing pre-eruptive Q-maps with the ribbons at around peak time of the flare of each eruption. Owing to noisy transverse field, and the discontinuous field distribution, the Q maps have patchy structure and we further apply a small threshold $|B_z|<15$~G to remove non-relevant QSLs. In addition, irrelevant QSLs are removed by applying a mask on the computed Q-maps.

The correspondence between the photospheric QSLs and flare ribbons has been shown in several studies. Theoretical studies predict that the flare ribbons are the photospheric foot prints of QSLs that encloses a twisted MFR \citep{demoulin1996, janvier2015}. The extremities of the ribbons are hook shaped for weakly twisted MFRs and are spiral shaped for highly twisted MFRs, as depicted in the cartoon of the 3D-model for eruptive flares in Figure~\ref{fig_cartoon}. Moreover, the characteristic 3D shape of QSLs associated with a twisted MFR depends on the height of the horizontal cut plane \citep{titov2007,savcheva2012c, Savcheva2015,zhaoj2016}. The QSL is S-shaped if the horizontal cut plane is at HFT and the cut planes lies below HFT, the QSL will appear as 2 (inverse) J-shaped with straight parts of J parallel to each other along the PIL. From this point of view of the MFR topology, the panels in Figure~\ref{fig_qf_11429} for the CME events from AR 11429, the QSLs are broadly spread over the flare ribbons especially the hook sections of the QSLs are in good morphological agreement. Such a remarkable match is an indication of the NLFFF models being capable of capturing hooked shape QSLs co-spatial with the observed flare ribbons, and was first reported by \citet{zhaoj2016} in the topological analysis of AR 12158 employing the Grad-Rubin based NLFFF model. 

From the above described MFR topology, the QSLs in the AR 12371 are also qualitatively match with the observed flare ribbons. The inverse-S ribbon exactly falls in the region outlined by QSLs of large Q-values. In fact, these QSLs can be regarded as a combination of 2 inverse-J shaped with a significant separation distance between the straight parts along PIL. We interpret the observed shape of the QSLs in the AR 12371 due to large part of sheared arcade surrounding the PIL, embedding the small-scale MFR. The sheared arcade becomes a large-scale twisted MFR only during onset of eruption by tether-cutting reconnection. The process of sheared arcade becoming large scale MFR is a dynamic process and may not be captured in vector magnetograms. Therefore, the static extrapolation results in a configuration with QSLs outlining the sheared arcade. Then the QSLs represents the boundaries of two connectivity domains separating the sheared arcade and the surrounding potential arcade. 

Another factor that could also contribute to capture the observed shape of QSL is the amount of twist in the magnetic configuration \citep{Pariat2012,savcheva2012a}. A more twisted MFR will present a more pronounced hook such that the it appears as spiral in shape. In our cases, the separation distance between the straight parts of QSL J-sections and their hook shape are very likely due to insufficient or weak twist. Moreover, the twist of these field lines increases as a consequence of the eruption. Meeting this twist criteria, in particular during the dynamic phase of the eruption, the QSL foot prints would be co-spatial precisely with the observed flare ribbons. 

\section{Summary and Discussion}
\label{sec_Summ}

Topological study of magnetic connectivity gradients provides more insights on the relation of magnetic structure of the sigmoid and flare ribbons in the frame work of standard model of eruptions. We studied the pre-eruptive magnetic structure of five CMEs launched from two successively erupting ARs. The modelled magnetic structure is largely resembles an inverse-S sigmoid in good agreement with the coronal plasma tracers in EUV observations. In all the eruption cases, the QSLs of large Q values are continuous enclosing core field bipolar region in which inverse-S shaped flare ribbon is observed. These QSLs essentially represent the large connectivity gradients between the  domains of twisted core flux within the inner bipolar region and the surrounding potential like arcade outside. It is consistent with the observed field structure largely with the sheared arcade. 

The QSL maps in the chromosphere are compared with the flare ribbons observed at the peak time of the flares. The flare ribbons are largely with inverse-S shape morphology with their continuity of visibility is missing in the observations. For the CMEs in the AR 12371, the QSLs outline the flare ribbons as a combination of two inverse J-shape sections but their straight sections being separated at the middle of the PIL. These QSLs are typically associated with the weakly twisted flux rope topology. Similarly, for the CMEs in the AR 11429, the QSLs are co-spatial with the flare ribbons both in the middle of the PIL and in the hook sections. This overall match of the observed flare ribbons with the photospheric QSLs is an indication that the NLFFF model of optimization approach reproduces the pre-eruptive magnetic structure to a very good extent. Earlier Grad-Rubin implementation of NLFFF model was reported to capture the hook shaped QSLs co-spatial with the flare ribbons \citep{zhaoj2016}.

However, we can notice that the co-spatiality is not precise to the predicted extent in the theoretical models of the MFRs. In the tether-cutting scenario, the MFR forms only during eruption dynamically at which time we expect hook shaped, two inverse-J sections of QSLs that are co-spatial with the observed flare ribbons \citep{savcheva2012b, janvier2015, zhaoj2016}. As per the view that the sheared arcades are weakly twisted MFRs in the sense that the magnetic field is dominated by the axial component \citep{Mackay2010}, we can regard the observed QSL shape in the AR 12371 as the combination of two inverse-J sections with significant separation distance between the straight parts along the PIL. Therefore,  the QSLs outlining the flare ribbons in the AR 12371 and less compact QSL hooks in the AR 11429 are likely due to the weakly twisted flux rope system. Additionally, difficulties related to input observations, construction of actual magnetic structure with the NLFFF modeling could also contribute to this discrepancy in the expected inverse-S (or two J-shaped ones) QSLs exactly co-spatial with the observed flare ribbons.

\section*{Permission to Reuse and Copyright}
Figures, tables, and images will be published under a Creative Commons CC-BY license and permission must be obtained for use of copyrighted material from other sources (including re-published/adapted/modified/partial figures and images from the internet). It is the responsibility of the authors to acquire the licenses, to follow any citation instructions requested by third-party rights holders, and cover any supplementary charges.



\section*{Conflict of Interest Statement}
The author declares that the research was conducted in the absence of any commercial or financial relationships that could be construed as a potential conflict of interest.



\section*{Acknowledgments} 
SDO is a mission of NASA's Living With a Star Program. The NLFFF code was developed by Dr. T. Wiegelmann of Max Planck Institute for Solar System. 3D field line rendering is due to VAPOR (\url{www.vapor.ucar.edu}) software. We acknowledge an extensive usage of the multi-node, multi-processor high performance computing facility at the Indian Institute of Astrophysics. I thank the reviewers for their generous comments and suggestions which improved the clarity of the text. 

\section*{Data Availability Statement}

The data sets used in this work is obtained from the instruments on board SDO spacecraft and is publicly available through \href{http://jsoc.stanford.edu/}{Joint Science Operations Center}.

\bibliographystyle{frontiersinSCNS_ENG_HUMS} 




\end{document}